\documentclass{elsart}

 \usepackage{epsfig}

\usepackage{amssymb}

\begin{document}

\begin{frontmatter}



\title{Proximity Networks and  Epidemics}

\author[label1,label2]{Zolt\'an Toroczkai\corauthref{cor1}},
\corauth[cor1]{Corresponding author. Tel.: +001 574 631 2618}
\ead{z.t@nd.edu}
\ead[url]{http://cnls.lanl.gov/$\sim$toro}
\author[label2]{Hasan Guclu}
\address[label1]{Department of Physics, University of Notre Dame, 225 Nieuwland Science
Hall, Notre Dame, IN, 46556}

\address[label2]{Center for Nonlinear Studies, Theoretical Division, MS B258, Los Alamos
National Laboratory, Los Alamos, NM, 87545}

\begin{abstract}
Disease spread in most biological populations requires the proximity of agents. 
In populations where the individuals have spatial mobility, the contact graph is generated 
by the ``collision dynamics" of the  agents, and thus the evolution of epidemics couples
directly to the spatial dynamics of the population. We first briefly review the properties and
the methodology of an agent-based simulation (EPISIMS) to model disease spread in 
realistic urban dynamic contact networks. Using the data generated by this simulation,
we introduce the notion of dynamic proximity networks which takes into account
the relevant time scales for disease spread: contact duration, infectivity period and rate of
contact creation. This approach promises to be a good candidate for a
unified treatment of  epidemic types that are driven by agent collision dynamics.  In
particular, using a simple model, we show that it can 
can account for the observed qualitative differences between the degree distributions of
contact graphs of diseases with short infectivity period
(such as air-transmitted diseases) or long infectivity periods ( such as HIV). 
\end{abstract}

\begin{keyword}
epidemics \sep social networks \sep spatial dynamics \sep agent-based modeling \sep scale-free 
network

\PACS 87.23.Cc \sep 87.23.Ge \sep 89.75.-k
\end{keyword}
\end{frontmatter}

\newpage
\section{Introduction}\label{intro}

Epidemics is the disease of the crowds. It is the process where a certain state of an
individual is transferred to other individuals by transport through a medium connecting
the agents such that eventually a finite fraction of the total population possesses that
state. In particular, if that state is an infectious illness, epidemics can have major 
negative consequences on a population, and thus the development of prevention and 
mitigation methods gain a crucial importance. In order to develop efficient strategies for
prevention and mitigation, however, one must understand the process of disease spread 
for the particular population in question. There has been considerable work devoted in the
past to the so-called compartmentalized models \cite{M04, E05, BC01, AM92} where the 
individuals are assigned one
of the finite number of compartments corresponding to their health state (such as 
Susceptible, Infected, and Recovered/Removed) combined with a uniform mixing model for the
disease transfer process for the individuals within each compartment. The fate of epidemics
in this framework is described by a set of coupled ordinary differential equations. While
such an approach is capable of producing reliable high level predictions for some diseases
(such as influenza) it cannot provide detailed information about non-aggregate variables,
which is crucial for developing efficient targeted vaccination and quarantine strategies. 
Most recently, however,  there has been  considerable effort invested 
\cite{CHEC03, Eubank04, BEM06, RF06, CBBV06} in agent-based or individual based 
approaches
which  build in-silico microscopic models of the population along with its dynamics. 
After a statistical validation with real data, the agent-based framework is then used as a test-bed for
 a number of different scenarios for epidemics spread and some of the possible
vaccination and quarantine strategies. Although this can be a useful tool for aiding decision
making, it is much less amenable to theoretical analysis, the simulation itself being a complex
system on its own right. Essentially, an agent-based model is a learning system \cite{G06}
whose structure and parameters are set such that it reproduces the statistics of real world data. 
Through the generalizing power of this system then one hopes to gain reliable insight into
data-scarce regions of the phase space.
Since the level of detail in agent-based modeling can be arbitrarily high,
this approach, however,  has the promise of giving specific, high resolution answers to questions like: ``Does vaccinating teachers and children between ages 2 and 12 have more impact on 
slowing disease spread than vaccinating cashiers? Which buildings should be closed in 
order to stop disease spread?", etc.  To understand the sensitivities in an agent-system for
disease spread, and perhaps draw some more general conclusions, one has to build  minimalist models for analyzing the data produced by such large-scale simulations. 
This paper presents a simple framework for understanding some of the topological features
of dynamic disease contact graphs, using data produced by a particular agent-based simulation,
EPISIM \cite{CHEC03, Eubank04, BEM06, BEMHSW04}, developed at Los Alamos National 
Laboratory (LANL).

In this article we shall confine ourselves to the case where disease is transferred through a
{\em contact process} between two individuals. Here contact is understood in a fairly loose
sense, only requiring that the two individuals be in the spatial proximity of each other. The
proximity distance required for disease spread is certainly disease dependent, ranging
from actual physical contact (such as in the case of sexually transmitted diseases) through 
a couple of feet to confinement in a building with common ventilation system (airborne diseases).

Another aspect that we will be considering is the {\em mobility} of agents. In contrast with, 
for example, the spread of viruses on computer networks which can be considered as a flow
process on a static structure, most populations are composed of mobile agents. As a result,
the contact network resulting from the ``collision dynamics'' of the agents is itself a dynamic 
entity. This is especially an important aspect for human populations in dense urban areas. 
 Currently, urbanization is in
an explosive stage \cite{NG}: the number of megacities (with over
ten million habitants) is estimated to increase from 14 in 1995
to 21 cities by 2015.  By 2030 it is estimated \cite{NG} that
over 60 percent of the world's population will live in cities.
For example, the population of S\~{a}o Paolo (Brazil),
the world's third most populated city, has grown from a population
of 265,000 to 18 million in the last 100 years. Almost half of S\~{a}o
Paolo's habitants were not born there \cite{NG}. Large cities act like
``magnets'' for people living in rural areas or smaller cities (especially
true in Third World countries), since over half of the gross domestic
product in most countries is made of industrial and commercial activities
taking place in these cities \cite{NG}, and thus they represent hopes for
prosperity.  A recent mathematical model
of aggregation accounting for this effect, was introduced by Leyvraz and
Redner \cite{LR02}.  Under such circumstances the problem of disease spread,
due to the dense nature of the contact fabric among people in cities becomes of real concern.

In the following we present a brief description of agent-based modeling using  EPISIM
as an example.  We recall some of the topological properties of the contact network obtained by
this simulation for the case of Portland, OR.  The key observation that we will be addressing 
in this paper refers to the
connectivity distribution of the people-people contact graph which seems to be rather different
from other measurements of contact graphs such as the sexual contact network measured by 
Liljeros et. al. \cite{LEASA01, AM88}. We will then propose a framework that can 
account for these differences in a unifying manner. We conclude by discussions on the limitation 
of the  model and possible extensions.

\section{An agent-based approach to epidemics}\label{ab}

The Transportation Analysis and Simulation System (TRANSIMS) 
\cite{TRANSIMS, N96, B00, BJM01, BBJKM02} developed  at LANL is an 
agent-based, cellular-automata model of traffic in a particular urban area (the first model 
was for Portland, OR, USA).  TRANSIMS decomposes the 
transportation planning task using three different time scales. 
A large time-scale associated with land use and demographic distribution 
is employed to create activities for travelers (activity 
categories such as work, shopping, entertainment, school, etc.).  Activity 
information typically consists of requests that travelers be at a certain 
location at a specified time, and includes information on travel modes 
available to the traveler. 
This is achieved by creating a synthetic population and endowing it 
with demographics matching the joint distributions given in the U.S. census data. 
The synthetic households are built by also using survey data from several
 thousands of households, which are observations made on the daily 
 activity patterns of each individual in the household. These activity patterns 
 are associated with synthetic households with similar demographics.  The 
 locations are estimated taking into account observed 
 land use patterns, travel times and transportation costs.  The 
 intermediate time-scale assigns routes and trip-chains to satisfy
  the activity requests. This is done by feeding the estimated locations into a routing 
  algorithm to find minimum cost paths through the transportation infrastructure 
  consistent with constraints on mode choice \cite{BJM01, BBJKM02}. 
   The third and shortest time-scale is associated with the actual execution of trip 
  plans in the road network.  This is done by a cellular automata simulation 
  \cite{N96,B00}
   through a detailed representation of the urban transportation network.
    The simulation resolves the traffic induced when everyone 
    tries to execute their plans simultaneously resolving distances 
    down to 7.5 meters and times down to 1 second. It provides an updated 
    estimate of time-dependent travel times for each edge in the network, including 
    the effects of congestion, which it feeds then to the router and location estimation 
    algorithms, which produce new plans. This feedback process continues iteratively 
    until it converges to a Òquasi - steady stateÓ in which no one can find a better path 
    in the context of everyone else's decisions. The resulting traffic patterns compare 
    well to observed traffic. The entire process estimates the demand on a transportation 
    network using census data, land usage data, and activity surveys.  More information 
    and including availability of the software can be obtained from Ref. \cite{TRANSIMS}.
    
EPISIM \cite{CHEC03, Eubank04, BEM06, BEMHSW04} is actually  one of the applications 
of TRANSIM and it is built on top of that. 
Diseases such as colds, flu, smallpox or SARS, are transmitted through air between
 two agents, if they spend long enough time in the proximity of each other, or in 
 building with closed air ventilation. This means, that we can assume that the 
 majority of the infections will take place in locations, like offices, shopping malls, 
 entertainment centers, mass transit units (metros, trams, etc.). Thus, by tracking 
 the people in our TRANSIMS virtual city, we can generate a bipartite contact 
 network, or graph, formed by two types of nodes, namely people nodes and
  locations nodes.  In the case of Portland, there are about 1.6 million people 
  nodes and 181,000 location nodes and over 6 million edges between them. 
  These are huge graphs, representing considerable challenges for the measurement 
  of their properties. 
  \begin{figure}[htbp]
\protect\vspace*{-0.1cm}
\epsfxsize = 3.4 in
\centerline{\epsfbox{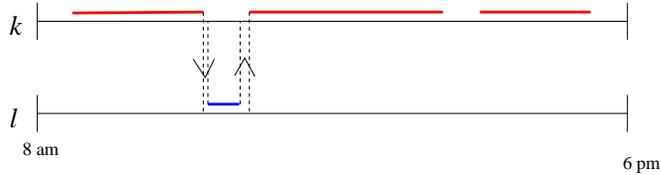}}
\protect\vspace*{-0.2cm}
\caption{An example for how directed edges are defined in the
location-location graph: $k$ and $l$ denote two locations, $k$ is an office while
$l$ is a nearby caffe. The horizontal axis is time and the thick lines denote
the presence of a person $p$ at that location. This diagram could stand for:
$p$ was at $k$ during mid-morning hours, then it went to $l$ for lunch, then
back to $k$ and then somewhere else (to doctor's appointment) then back to $k$ then
somewhere else, etc. } \label{sketchll}
\end{figure}

 Let us denote by $L$ the set of locations and by $P$ the set of people.
The people and locations are indexed by integers, called {\em person-id}
and {\em location-id}. The vertex set of a graph $G$ will be referred to
as $V(G)$ while the edge set will be referred to as $E(G)$. The degree
of a vertex $v \in V(G)$  is the number of edges incident on $v$.
The activities in the EpiSIMs graphs have a time-periodic character. However, on
average, people typically resume their activity patterns after 24 hours and thus 
graphs corresponding to different week-days a similar to one another.
The time labels to be defined below are measured time intervals for a duration
of 24 hours and the graphs defined also refer to this 24 hour period.
The temporally resolved bi-partite graph of people
and locations is denoted as $(G_{PL},\beta)$, where the only edges
present are between individuals and the locations they visit. The vertex set
is defined as $V(G_{PL})=P\cup L$. An
edge $e \in E(G_{PL})$ is defined by the ordered pair $(p,l)$ where $p\in P$
is the person-id of the individual and $l\in L$ is the location-id of the location
which it visited.
$\beta(e)$ signifies a time label associated with the  edge $e \in E(G_{PL})$, and it is
defined as the set of non-overlapping time intervals
$\beta(e) = \beta(p,l)=\{ I^{(1)}(e), I^{(2)}(e), ... \}$, given by $I^{(j)}(e)=
I^{(j)}(p,l)=
[t_{in}^{(j)}(e),t_{out}^{(j)}(e)]$
between the
``in-time" $t_{in}^{(j)}$ and  ``out-time'' $t_{out}^{(j)}$ to and from $l$ of $p$. The reason
for a number of different time intervals is that the same person can visit several times
the same location during a day (such as office - lunch - office, etc.).
If two intervals $I(e)$ and $I(e')$ are non-overlapping, then we define $I(e) <
I(e')$, iff $t_{out}(e) \leq t_{in}(e')$. We consider two other types of contact
networks: the \emph{people-people} graph, denoted by $(G_P,\pi)$,
and the \emph{location-location} graph, $(G_L,\lambda)$.
In $G_P$, an individual $u\in P$ is represented by one vertex. There is an edge
$e=(u,v)\in E(G_P)$ if the individuals $u,v \in P$ have come  into contact, i.e.,
if $\exists$ $e_u,e_v \in E(G_{PL})$ and $l \in L$, such that $e_u = (u,l)$, $e_v = (v,l)$
and $\beta(e_u) \cap \beta(e_v) \neq 0$. The time label associated with this
edge therefore is calculated by $\pi(e) = \bigcup_{l\in L} \beta(e_u) \cap
\beta(e_v)$, composed of intervals of time when they have {\em shared} the
same location (any) during a day.

The location-location graph $(G_L,\lambda)$, is a {\em directed graph}, where
every vertex represents a location from $L$,
and a directed edge ${\mathbf e}\in E(G_L)$ is defined by the
{\em ordered pair} ${\mathbf e}=(k,l)$, $k,l\in L$, if there is at least one person
$p \in P$ going from $k$ to $l$ anytime during the day, see for an example Figure
\ref{sketchll}. Naturally, since a person can be in a single location at a
given time instant, $\beta(p,k) \cap
\beta(p,l) = 0$ ($k \neq l$). Thus ${\mathbf e}=(k,l)$ is an edge from $k$ to $l$, if $\exists$
$I^{(j)}(p,k) \in \beta(p,k)$ and $I^{(m)}(p,l) \in \beta(p,l)$ such that
$I^{(j)}(p,k) < I^{(m)}(p,l)$ and for $\forall$ $I(p,n) \in \beta(p,n)$ with $\forall$
$n \in L \setminus \{k,l\}$, either $I(p,n) < I^{(j)}(p,k)$ or $I^{(m)}(p,l) < I(p,n)$.
An entrance time of $p$ at $l$ coming from $k$, is obviously $t_{in}^{(m)}(p,l)$. Since this
is a continuous time, discrete event process, these entrance times into location $l$
of people coming directly from location $k$ during a day, can be ordered into a set
$\lambda(k,l) = \lambda({\mathbf e})$, which forms the {\em weight label} of edge
${\mathbf e}$ in $E(G_L)$.
We are interested in entrance times to a location because this way we are able to
record when an infection enters a location and thus the time after which possible infections
can occur for people visiting that location.
\begin{figure}[htbp]
\protect\vspace*{1.0cm} \epsfxsize = 4.5 in
\centerline{\epsfbox{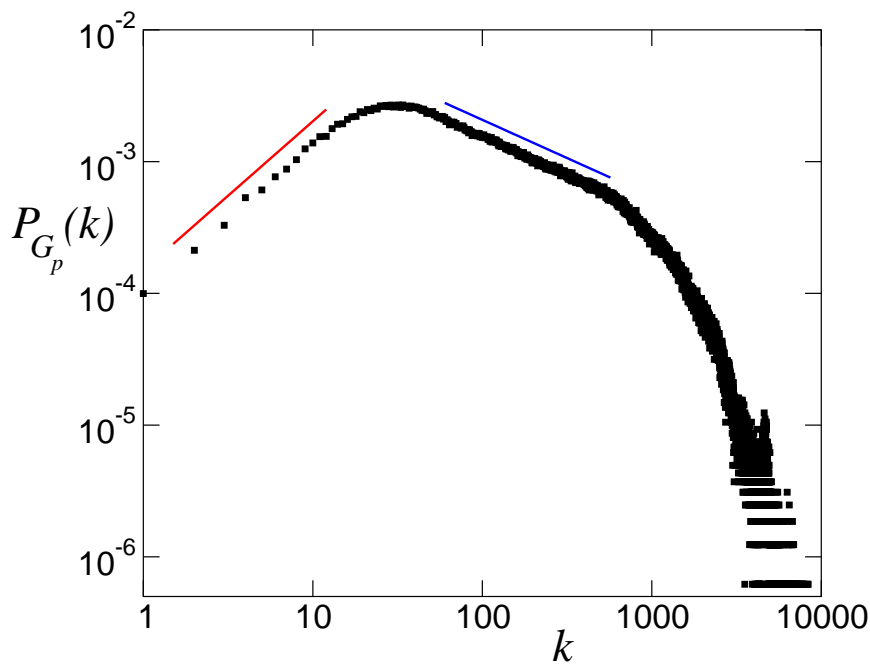}}
\protect\vspace*{-0.0cm}
\caption{Degree distribution of the people-people contact network $G_P$ (see text for definition) in 
EPISIM Oregon data. The measurement of the degree distribution is done one a single
instance of the $G_P$ (and thus $G_{PL}$) graph. The contact time threshold used was one hour. 
There are $5788$ points in the data set and $k_{max} = 8368$.
The slopes of the straight lines are $1.13$ and $-0.58$, respectively. }\label{fig2}
\end{figure}
Figure \ref{fig2} shows the measured degree distribution of the people-people contact graph $G_p$ 
keeping only those edges which come from time-stamp overlaps of at least one hour.
As one can see the degree distribution has an exponential cut-off (at about $k=700$) and a 
peak, not 
reminiscent of pure power-law (scale-free) networks.  Although the curve seems to have power-law behavior portions of it,
nothing really can be concluded based on this data, since this behavior is only about over one
decade. The exponential cut-off is a must, because an individual cannot be in contact with $O(N)$
other people during a day. 
The graph integration interpretation that we will present in the next section, however,
identifies some key ingredients that might be responsible  for the shape of this distribution.

\section{Proximity networks}\label{proxnets}

Some of the most  difficult questions in epidemics concern  the spatio-temporal dynamics,
as opposed to looking at disease spread as a percolation process on fixed structures. 
Ref. {\cite{V06}} is a most
recent attempt studying epidemics as a branching process building effective contact graphs
with scale-free and small-world behavior. This paper illustrates that the details of the dynamics can
have drastic effects on disease spread by creating contact graphs with heterogeneous 
structures.

As we have seen from the EpiSim example above, the true contact graph is a dynamic structure
with  temporal behavior that can be encoded as time-stamps associated with edges. If one
neglects the time-stamps, the graph obtained is the {\em maximum} contact graph, showing
which individuals came in contact at all, during a day. On the other hand, at every instant,
the graph of contacts is a set of disconnected small graph clusters showing which individuals
are in contact {\em right at that moment}. Naturally, the maximum contact graph is much denser than
the instantaneous contact graphs. 
Due to the mobility of the individuals, the contacts are changing,
and if one would like to know who was in contact with whom over a period of time, we need to 
{\em integrate} or {\em collapse} the instantaneous contact graphs in that time period. If our goal
is to produce an {\em effective} contact graph for a particular disease, so that the disease spread can
be treated as a flow process on this effective graph, we need to collapse the instantaneous contact graphs over the typical infectivity period of an individual. For the sake of brevity we shall refer to
the effective integrated contact graph as the {\em proximity network} of the disease. 
In case of aerial-born diseases like SARS
and smallpox the infectivity period is on the order of several days, while for some sexually transmitted
diseases, like HIV, it is much longer, and can be on the order of years. In the former case the
maximum degree of a node is relatively small (constant compared to the system size) 
bounded by the number of contacts an agent can possibly make
in a short period of time (days), whereas in the latter case the number of accumulated contacts
(and thus the number of possibly infected people) can be very large. In this latter case is when we
expect to have scale-free behavior for the degree distribution of the contact graph. Indeed, 
several measurements on the distribution of sexual interactions in human populations \cite{LEASA01, AM88} seems to confirm this expectation. In the remainder of this article we show a very simple
model for proximity networks which can capture some of these observations, in particular the
scale-free character.

Let us denote the instantaneous contact graph at instant $t$ by $G(t)$. The
time-integrated graph, or {\em proximity graph} ${\overline{G}(T)}$ is given by the
union of all edges of the instantaneous graphs from time $0$ to time $T$. In
other words, if $\mathbf{A}(t) = \{ a_{ij}(t)\}$ is the adjacency matrix of $G(t)$, the adjacency
matrix $\overline{\mathbf{A}}(T)$ of ${\overline{G}(T)}$ is given by:
\begin{equation}
\overline{a}_{ij}(T)=\bigvee_{t=0}^{t=T}a_{ij}(t)
\end{equation}
where $a\vee b = 0$ if and only if $a=0$ and $b=0$, otherwise $a\vee b = 1$, $a,b \in \{ 0, 1\}$.
The dynamics of the integrated network is determined by the dynamics
of the contact making process between pairs of agents $i$ and $j$. Depending on the
nature of the potentially disease transmitting contacts the matrix elements $a_{ij}(t)$ 
might be subject to ``exclusion constraints". For example, in the case of sexual contacts,
the instantaneous graphs are made of a collection of dimers and/or isolated nodes.
This induces the constraint $\sum_{j, j\neq i} a_{ij} \leq 1$, $i=1,...,N$ on the matrix elements
of $G(t)$.  For diseases spread by air, the transmission usually happens in locations which
have a limited capacity. Assuming that all agents within a location can be infected if at least 
one of them is infectious (for example due to shared ventilation system), the contact graph within a location is a clique.  The constraints imposed on the instantaneous adjacency matrix 
$\mathbf{A}(t)$ can be thus be formulated as:
\begin{eqnarray}
(1-a_{ik})a_{ij}a_{jk} = 0\;,\;\;\;\;\mbox{for all}\;\;\;\;i,j,k \in {1,...,N} \label{cq}\\
\sum_{j,j\neq i} a_{ij} \leq K\;,\;\;\;\;\mbox{for all}\;\;\;\;i \in {1,...,N} \label{K}
\end{eqnarray}
where $K$ is the maximum clique size (maximum location capacity).
The first of these equations is a necessary and sufficient condition for a graph to be
the disjoint union of cliques and the second limits the size of the cliques to $K$. Equation
(\ref{cq}) expresses the fact that all connected triplets must form a triangle (if $i$ and 
$j$ are connected and $j$ and $k$ are connected, then $i$ and $k$ are connected as well
- transitivity). In the physics network literature the alternative formulation of the same
condition is that the clustering coefficient of $G(t)$ is unitary. 

Here we will not consider a full theory of dynamic proximity networks, that
will be developed elsewhere. Instead, we introduce  the most simplistic model of 
network growth which still reproduces the qualitative features of the observations in the beginning
of this section.

The probability $\overline{p}_{ij}(T)$ that nodes $i$ and $j$ are connected in the 
proximity graph at time $T$ increases for larger values of $T$. Assuming a uniform 
link generation picture, the probability that in the next step a potentially disease spreading
connection/contact  takes place between agents $i$ and $j$ is written as 
$\rho g_i g_j$ where the weight $g_i$ quantifies the ``gregariousness'' of agent $i$,
its propensity to generate new links. Note that this model does not explicitly resolve the
exclusion constraints (\ref{cq})-(\ref{K}). One can think of the parameter $\rho$ effectively 
incorporating the spatial information, which should be resolved in spatial models for contact dynamics.
The probability that after $T$ steps nodes $i$ and $j$ are
connected is given by:
\begin{equation}
\overline{p}_{ij}(T) = 1-(1-\rho g_i g_j)^{T} \;,\;\;\;T \geq 1\;. \label{plink}
\end{equation}
The parenthesis in (\ref{plink}) represents the probability of nodes $i$ and 
$j$ {\em not} connecting in one
step, and its $T$-th power is for non-connection during all steps. One minus this probability
is obviously the connection probability during any of the steps $1,2,....,T$.
According to this, highly gregarious people will
more likely be connected to each other than less gregarious. 
They will also get connected earlier than others. $\rho$ is a parameter which makes
$\rho g_i g_j \leq 1$, but at this point is a free parameter. If we want to stay close to the
claim that in one step a node does not accumulate more links than it is allowed
by the exclusion conditions, we need to consider $\rho$ to be a small number.  
If all nodes have the same gregariousness parameter $g$ then we recover
a growing binomial random graph model and the degree distribution of the proximity graph will
always stay a Poisson distribution with a parameter that grows exponentially with time until
the graph becomes a complete graph. This is certainly expected, given that there is no heterogeneity
in the mixing among agents.

There are certainly many possible, more realistic extensions to 
our model, in particular making the gregariousness coefficients time dependent and taking
into account more explicitly the exclusion constraints. Time dependent gregariousness coefficients
would
correspond to cases where, for example, there is an increasing cost of adding a link (because
it involves traveling further away) or the health state of the agent (e.g., infected or not) can modify
their ability to generate new links. Here we will only study the case where these coefficients
are time-independent. 
\begin{figure}[htbp]
\protect\vspace*{-0.1cm} \epsfxsize = 5.4 in
\centerline{\epsfbox{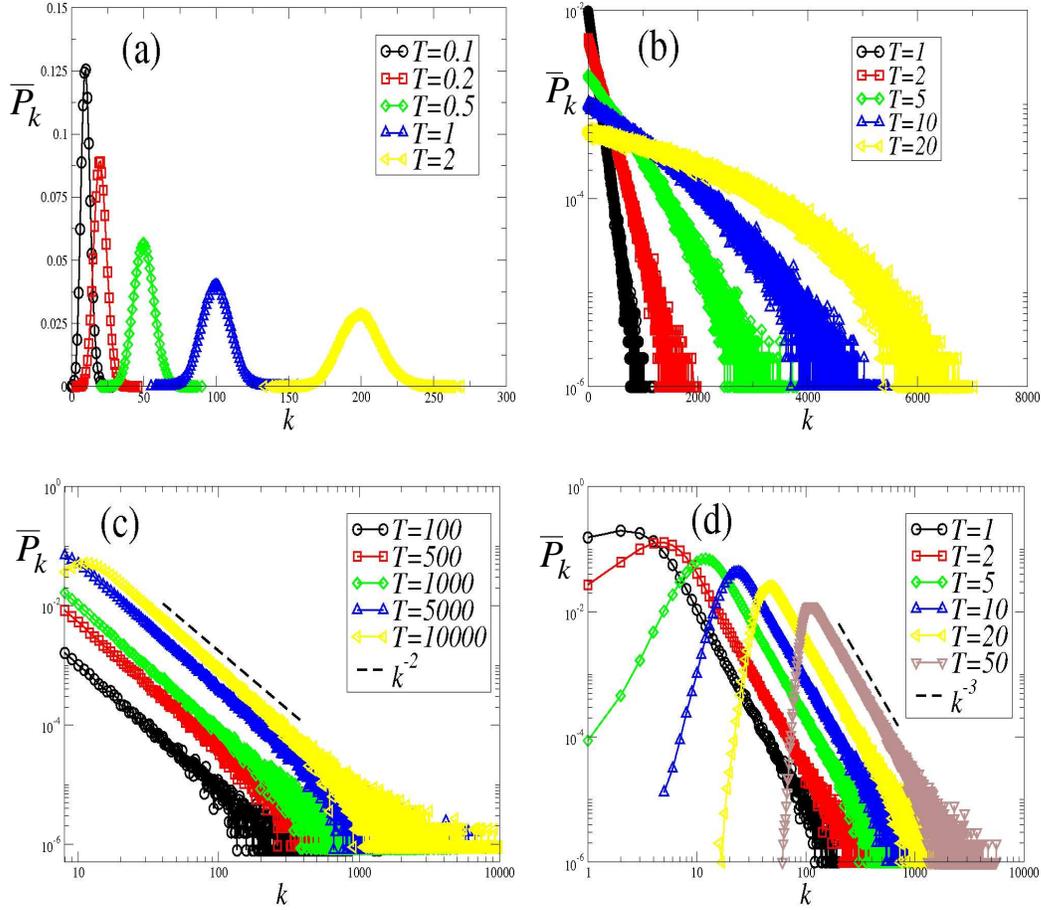}}
\protect\vspace*{-0.2cm}
\caption{Degree distributions of the proximity network for various distributions 
of the gregariousness. a) constant $g$, (b) exponential distribution for $g$, (c) and (d) power-law $g$.
All networks have $N=10^4$ nodes.  }\label{fig3}
\end{figure}

Under what conditions for the gregariousness distribution of a population will we observe 
power-law (scale-free) degree distributions for its proximity network? One expects that 
populations where all individuals have similar gregariousness values no power-law 
should be observed for the degree distribution of ${\overline{G}(T)}$, whereas heterogeneous
distributions for $g_i$ would likely generate distributions with a power-law regime in them.
The expected degree of node $i$ in ${\overline{G}(T)}$ is:
\begin{equation}
\overline{d}_i(T)=\sum_{j=1}^{N}\overline{p}_{ij}(T) =\sum_{j=1}^{N} \left[ 1-(1-\rho g_i g_j)^{T}\right]
\end{equation}
For small $\rho$, $\rho T \ll 1$, this is simply $\overline{d}_i(T) \simeq \rho T (\sum_{j} g_j) g_i$, i.e,
it is directly proportional to its gregariousness coefficient. This certainly makes sense in a social
context since more gregarious people will have on average more contacts than others. In this 
limit (small $\rho$, and not too large $T$ values, such that  $\rho T \ll 1$) our model is similar to the Chung Lu model \cite{CL02} of random power-law graphs with expected
degree sequences. The difference is that in our case the $g_i$-s are random variables drawn independently from a given distribution, while in the Chung Lu model the node weights 
are prescribed functions of their index. In the small $\rho$ limit initially the graph will 
form disconnected clusters, but it does not strictly obey the exclusion conditions. As time
goes on the links accumulate on the proximity network and one can obtain a 
regime where depending
on the gregariousness distribution, scale-free contact networks emerge. If the agents 
are not removed from the system, but keep accumulating contacts, eventually a finite
network will reach the complete graph limit and stop there. If the infectivity period is finite,
however, the network will reach a steady state structure characteristic to the population dynamics
and the disease. This might be scale-free, homogeneous, or anything in between. 

Going back to the EPISIM contact network data we notice that this simplistic model (see Fig. 3) reproduces
qualitatively some of the key features shown in Fig. \ref{fig2}. Although the EPISIM distribution has a sharp, exponential cut-off, it shows a 
tendency of forming a power-law tail before the cut-off  just as in our model. In addition, it seems that the low-$k$
tail shows a similar power-law tendency, also in qualitative accordance with the data generated 
by our simplistic model, see Fig. \ref{fig3}(d).  With this model we can also generate distributions (not shown) with small (including
sub-unitary) exponents for the power-law regime followed by a sharp cut-off, matching closer the 
distribution in Fig. \ref{fig2}.

\section{Conclusions}

We presented the basis of a framework to account for the dynamics of contacts 
in epidemic processes, through the notion of dynamic proximity graphs. 
By varying the integration time-parameter $T$, which is the period of infectivity one can give a simple account for some of the differences in the observed contact
networks for different diseases, such as smallpox, or AIDS.
Our simplistic model 
also seems to  shed some light on the shape of the degree distribution of the measured people-people 
contact network from the EPISIM data. 

We certainly do not claim that the simplistic graph integration model [Eq.~(\ref{plink})] above is a 
good model for dynamic contact graphs. It only contains the essential ingredients for such processes
to produce a qualitative agreement with some observations. We expect that further refinements and extensions  to this picture, in particular deriving the link-probabilities in the dynamic proximity
graph from more realistic contact dynamics should improve the agreement between models
and data.

\section*{Acknowledgements}

The authors thank for support from the National Nuclear Security
Administration of the U.S. Department of Energy at Los Alamos National
Laboratory under Contract No. DE-AC52-06NA25396, and S. Eubank,  for useful discussions
and availability of data.




\begin{thebibliography}{99}



\bibitem{M04} J.D. Murray, {\em Mathematical Biology II}, 3rd ed. Springer (2004).

\bibitem{E05} L. Edelstein-Keshtet, {Mathematical Models in Biology}, SIAM (2005).

\bibitem{BC01} F. Bauer, C. Castillo-Ch\'avez, {\em Mathematical Models in Population Biology 
and Epidemiology} Springer, (2001).

\bibitem{AM92} R.M. Anderson, R.M. May, {\em Infectious Diseases in Humans}, Oxford 
University Press, Oxford, (1992).

\bibitem{CHEC03} G. Chowell, J.M. Hyman, S. Eubank and C. Castillo-Ch\'avez, {\em Phys. Rev. E},
{\bf 68}, 066102 (2003).

\bibitem{Eubank04} S. Eubank, H. Guclu, V.S. Anil Kumar, M.V. Marathe, A. Srinivasan,
Z. Toroczkai and N. Wang, {\em Nature} {\bf 429}, 180 (2004).

\bibitem{BEM06} C. Barrett, S. Eubank, and M. Marathe, in {\em Interactive Computing: A New Paradigm}, eds. D. Goldin, S. Smolka, and P. Wegner,  Springer, Berlin (2006), pp. 353.

\bibitem{BEMHSW04} C. Barrett, S. Eubank, M. Marathe, H. Mortveit, A. Srinivasan, and N. Wang, 
{\em Proc. 15th annual ACM-SIAM Symposium on Discrete Algorithms, New Orleans},  718 (2004).

\bibitem{RF06} S. Riley, N.M. Ferguson, {\em Proc. Natl. Acad. Sci. USA} {\bf 103}, 12637 (2006).

\bibitem{CBBV06} V. Colizza, A. Barrat, M. Barthelemy and A. Vespignani, 
{\em Proc. Natl. Acad. Sci. USA} {\bf 103}, 2015 (2006). 

\bibitem{G06} N. Gilbert et.al. {\em Journal of Artificial Societies and Social Simulation} {\bf 9}(2) (2006). 

\bibitem{NG} E. Zwingle, 
{\em Natl. Geogr. Mag.} {\bf 202} 70-99 (2002).

\bibitem{LR02} F. Leyvraz, S. Redner,
{\em Phys. Rev. Lett. } {\bf 88} 068301 (2002).

\bibitem{LEASA01} F. Liljeros, C.R. Edling, L.A.N. Amaral, H.E. Stanley, {\em Nature}  {\bf 411},
907 (2001).

\bibitem{AM88} R.M. Anderson, R.M. May, {\em Nature} {\bf 333}, 514 (1988).

\bibitem{TRANSIMS} {\tt http://transims.tsasa.lanl.gov}, {\tt http://www.transims.net} 

\bibitem{N96} K. Nagel, {\em Phys. Rev. E} {\bf 53}, 4655 (1996).

\bibitem{B00} K. Nagel, P. Stretz, M. Pieck, S. Leckey, R. Donnelly,
and C.L. Barrett, Technical Report LA-UR-97-3530 (1997).

\bibitem{BJM01} C.L. Barrett, R. Jacob and M.V. Marathe, {\em SIAM J. Comp.} {\bf 30}, 809 (2001).

\bibitem{BBJKM02} C. Barrett, K. Bisset, R. Jacob, G. Konjevod, M. V. Marathe, in
{\em Proceedings of the European Symposium on Algorithms, Lecture Notes in Computer Science}, Springer, 126 ( 2002).

\bibitem{V06} A. Vazquez, {\em Phys. Rev. Lett.} {\bf 96}, 038702 (2006).

\bibitem{CL02} F. Chung Graham, L. Lu, {\em Annals of Combinatorics } {\bf 6}, 125 (2002). 




\end{thebibliography}
\end{document}